\documentclass[12pt]{article}
\usepackage{graphicx}
\usepackage{float}
\newcommand{\bb}{\begin{equation}}
\newcommand{\ee}{\end{equation}}
\newcommand{\ba}{\begin{eqnarray}}
\newcommand{\ea}{\end{eqnarray}}

\begin{document}

\title{{\bf Two-Point Pad\'{e} Approximants for the Deflection of Light in the Schwarzschild Black Hole Metric}}

\author{
Don N. Page
\thanks{Internet address:
profdonpage@gmail.com}
\\
Department of Physics\\
4-183 CCIS\\
University of Alberta\\
Edmonton, Alberta T6G 2E1\\
Canada
}

\date{2026 April 21}

\maketitle
\large
\begin{abstract}
\baselineskip 20 pt

The deflection angle of a light ray passing the Schwarzschild (spherically symmetric vacuum) black hole was calculated by Charles Galton Darwin in 1959 in terms of the elliptic integral of the first kind.  This calculation has been repeated many times and has also been given approximately in terms of elementary functions for impact parameters that either are not too small or are close to the critical impact parameter.  Here I present Pad\'{e} 2-point approximants of order [2,2] (quadratic numerators and denominators), relating the critical impact parameter divided by the actual impact parameter to the exponential of the negative of the deflection angle, that fairly accurately cover the full range of impact parameters greater than the critical impact parameter, which is the case for all photon trajectories that remain outside the black hole.  I also present a simpler quadratic approximation that works as well in the middle of the range but not so well at the extremes.

\end{abstract}

\normalsize

\baselineskip 22 pt

\newpage

\section{Introduction:  Charles G.\ Darwin's Calculation}

The observational confirmation, during the 1919 total eclipse of the sun, of Einstein's prediction of twice the Newtonian deflection of light, became a world-famous confirmation of Einstein's theory of general relativity.  The factor of two is very accurate for the weak gravitational field of the sun, but it became of interest to calculate the exact deflection, especially for the strong gravitational field of a black hole.  This was done for the nonrotating Schwarzschild metric (the only case considered here) by Charles Galton Darwin (the grandson of Charles Robert Darwin) in 1959 \cite{Darwin}.

Here I shall use classical general relativistic length units with $G = c = 1$ for a spherically symmetric static vacuum gravitational field of mass $M = GM/c^2$ and set Darwin's `perihelion' distance (he takes the gravitating mass to be the sun with mass $m$) to be
\bb
P = \frac{2M}{v},
\label{P}
\ee
in terms of the dimensionless parameter I shall call $v$, and Darwin's parameter
\bb
Q = \sqrt{(P-2M)(P+6M)} = \frac{2Mq}{v}
\label{Q}
\ee
in terms also of the dependent dimensionless parameter I shall call $q$,
\bb
q = \sqrt{(1-v)(1+3v)}.
\label{q}
\ee
Darwin's formula for the deflection angle $\mu$ uses the incomplete elliptic integral of the first kind,
\bb
F(\phi|m=k^2) = F(k,\phi) 
= \int_0^\phi \frac{d\varphi}{\sqrt{1-m\sin^2{\varphi}}}
= \int_0^\phi \frac{d\varphi}{\sqrt{1-k^2\sin^2{\varphi}}},
\label{Felliptic}
\ee
with the elliptic integral {\it parameter}
\bb
m \equiv k^2 = \frac{Q-P+6M}{2Q} = \frac{q-1+3v}{2q}.
\label{m}
\ee
(Note that Darwin uses $m$ for the gravitational mass rather than for the elliptic integral {\it parameter} as I do, but instead he uses the 
{\it elliptic modulus} or {\it eccentricity}
\bb
k \equiv \sqrt{m} = \sqrt{\frac{Q-P+6M}{2Q}} 
= \sqrt{\frac{q-1+3v}{2q}}.)
\label{k}
\ee

In terms of the mass $M$ and my dimensionless parameter $v$ for the inverse perihelion mass in units of the Schwarzschild radius $R_S = 2M$, the impact parameter (which Darwin labels by $\ell$) is what I label as $b$,
\bb
\ell = b = \frac{2M}{v\sqrt{1-v}} = \frac{3\sqrt{3}M}{\beta}
\label{b}
\ee
in terms of my key dimensionless parameter $\beta$,
\bb
\beta \equiv \frac{3\sqrt{3}M}{b} = \frac{3\sqrt{3}}{2}v\sqrt{1-v},
\label{beta}
\ee
which varies from $\beta = 0$ for infinite impact parameter $b$ to 
the critical (maximum) value of $\beta = \beta_c = 1$ for the critical impact parameter $b_c = 3\sqrt{3}M$ that leads to the circular photon orbits at the critical perihelion radius $P_c = r_c=3M$ and the critical value of $v = 2M/P$ as $v_c = 2M/P_c = 2M/r_c = 2M/(3M) = 2/3$.

The {\it amplitude} for the upper limit of the elliptic integral $F(\phi|m)$, which Darwin labels as $\phi_1$, I shall call simply $\phi$, and Darwin gives the square of its sine as
\bb
\sin^2{\phi} = \frac{2Q}{3P-6M+Q} = \frac{2q}{3-3v+q}.
\label{sinphi}
\ee

In terms of these dimensionless parameters $v$, $q$, $m=k^2$, and $\phi$, Darwin's formula for the deflection angle $\mu$ is
\bb
\mu = \frac{4}{\sqrt{q}}F(k,\phi)-\pi 
    = \frac{4}{\sqrt{q}}\int_0^\phi 
    \frac{d\varphi}{\sqrt{1-m\sin^2{\varphi}}}-\pi.
\label{mu}
\ee

I then define my second key dimensionless parameter $\alpha$ as the exponential of the negative of the deflection angle $\mu$:
\bb
\alpha \equiv e^{-\mu}
= \exp{\left(\pi - \frac{4}{\sqrt{q}}F(k,\phi)\right)}.
\label{alpha}
\ee

Like my key dimensionless parameter $\beta$ that directly determines the impact parameter $b = 3\sqrt{3}M/\beta$ (once the gravitating mass $M$ is known to set the length scale), my second key dimensionless parameter $\alpha = e^{-\mu}$ directly determines the light deflection angle $\mu = -\ln{\alpha}$, and $\alpha$ also ranges from 0 (for $\beta = 1$ that gives infinite deflection angle $\mu$) to 1 (for $\beta = 0$ that gives infinite impact parameter $b = 3\sqrt{3}M/\beta = \infty$ and zero deflection angle, $\mu = 0$, and hence $\alpha = e^{-\mu} = 1$).

\begin{figure}
\includegraphics[width=1.0\columnwidth]{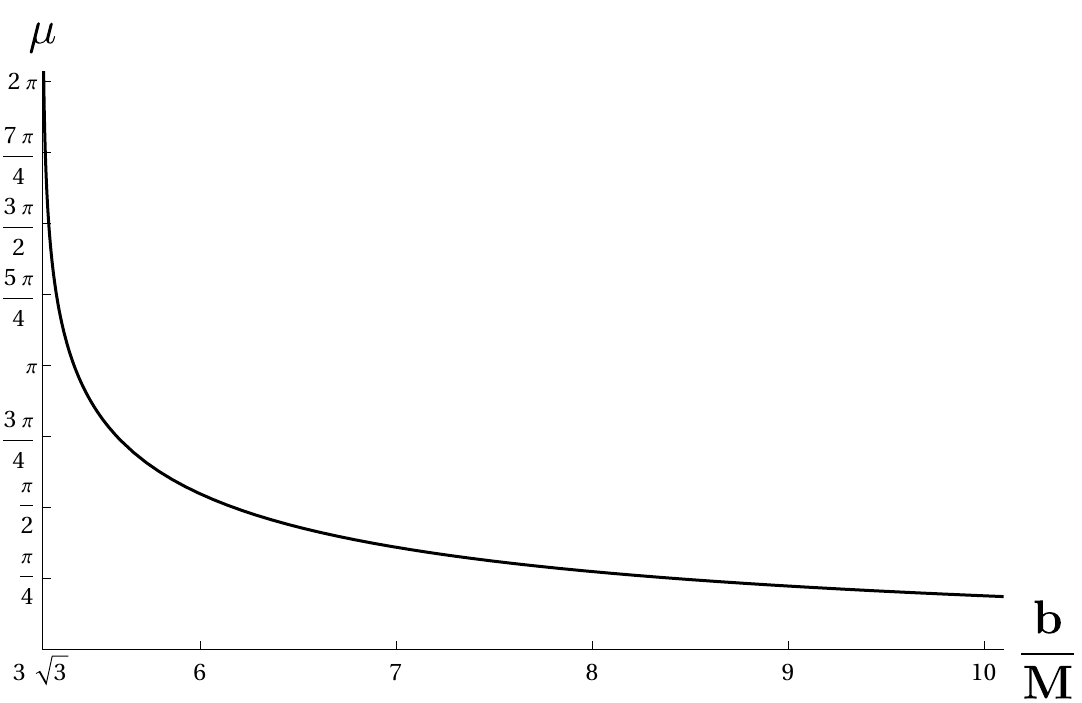}
\caption{This graph gives the `exact' (within the numerical precision of Mathematica) light deflection angle $\mu \equiv  -\ln{\alpha} = \frac{4}{\sqrt{q}}F(k,\phi)-\pi$ as a function of $\frac{b}{M} \equiv \frac{3\sqrt{3}}{\beta}$.
The graph is plotted at the left end up to just past a 360-degree deflection angle, $\mu = 2\pi$, which corresponds to $\alpha = e^{-2\pi} \approx 0.001\,867\,443$, $\beta \equiv \frac{3\sqrt{3}M}{b} \approx 0.998\,745$, and $\frac{b}{M} \approx 5.202\,680 \approx 3\sqrt{3} - 0.006\,528$, with this last number (the difference from the critical impact parameter value) too small to be distinguished from 0 in the graph.  At this scale for the graph, one could also not see the difference between the `exact' $\mu(b/M)$ plotted and either the Pad\'{e} approximant $\mu_p$ or the simpler quadratic approximation $\mu_q$ to be developed later and whose small errors will be shown on a graph with a much larger scale.}
\end{figure}

\begin{figure}
\includegraphics[width=1.0\columnwidth]{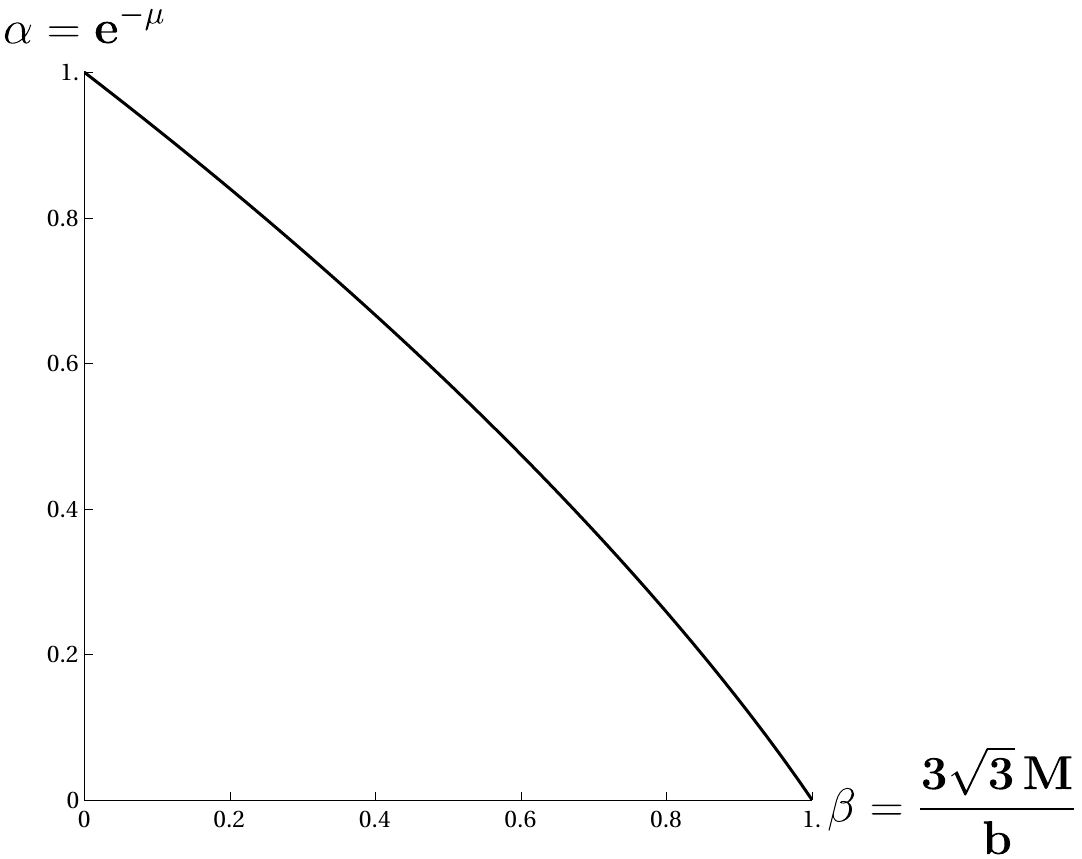}
\caption{This graph gives the `exact' (within the numerical precision of Mathematica) function 
$\alpha = e^{-\mu} = \exp{\left(\pi - \frac{4}{\sqrt{q}}F(k,\phi)\right)}$ of the light deflection angle $\mu \equiv  -\ln{\alpha} = \frac{4}{\sqrt{q}}F(k,\phi)-\pi$ as a function of $\beta \equiv \frac{3\sqrt{3}M}{b}$ that is the ratio $\frac{b_c}{b}$ of the critical impact parameter $b_c = 3\sqrt{3}M$ to the actual impact parameter $b$.  Note that at the left end of the graph, where the impact parameter $b$ is taken to infinity, $\beta = \frac{b_c}{b}$ goes to 0, the deflection angle $\mu$ also goes to 0, and $\alpha = e^{-\mu}$ goes to 1.  At the right end, where the impact parameter $b$ is taken to the critical value $b_c = 1$, the deflection angle $\mu$ goes to infinity, and $\alpha = e^{-\mu}$ goes to 0.  At this scale for the graph, one could not see the difference between the `exact' $\alpha(\beta)$ plotted and either the Pad\'{e} approximant $\alpha_p(\beta)$ or the simpler quadratic approximation $\alpha_q(\beta)$ to be developed later and whose small errors will be shown on a graph with a much larger scale.}
\end{figure}

\section{Asymptotic Behavior to Be Fit}

Here I wish first to give order-[2,2] two-point Pad\'{e} approximants for $\beta(\alpha)$ and for $\alpha(\beta)$, namely
\bb
\beta(\alpha) \approx \beta_p(\alpha) 
= \frac{1 + A\alpha - (1 + A)\alpha^2}{1 + B\alpha + C\alpha^2},
\label{betaa}
\ee
and
\bb
\alpha(\beta) \approx \alpha_p(\beta) 
= \frac{1 + F\beta - (1 + F)\beta^2}{1 + G\beta + H\beta^2},
\label{alphaa},
\ee
with the subscript $p$ denoting ``Pad\'{e},'' where the forms of the expressions ae chosen so that $\beta_p(0)=1$, $\beta_p(1)=0$, $\alpha_p(0)=1$, and $\alpha_p(1) = 0$, and the constants $A$, $B$, $C$, $F$ (here a constant rather than the elliptical integral), $G$, and $H$ are determined and given explicitly below by the following three requirements for each triplet:

(1)  Albert Einstein \cite{Einstein:1916vd} famously showed that for an impact parameter $b \gg M$ ($\beta = 3\sqrt{3}M/b \ll 1$), general relativity gives the deflection of light as 
\bb
\mu = -\ln{\alpha} \approx \frac{4M}{b} = \frac{4}{3\sqrt{3}}\beta.
\label{muverysmallbeta}
\ee

(2)  Reuben Epstein and Irwin I. Shapiro \cite{Epstein:1980dw},
Ephraim Fischbach and Belvin S.\ Freeman \cite{Fischbach:1980su}, and many others,
such as Gary W.\ Richter and Richard A.\ Matzner \cite{Richter:1982zz},
have calculated that to second order in $M/b$, the deflection angle of light is
\bb
\mu = \frac{4M}{b}\left[1 + \frac{15\pi}{16}\frac{M}{b}\right]
+O\left[\left(\frac{M}{b}\right)^3\right],
\label{musmallbeta}
\ee
which gives
\bb
\beta = x\mu(1-y\mu)+O(\mu^3) = x(1-\alpha)[1-(y-1/2)(1-\alpha)]+O[(1-\alpha)^3]
\label{betasmallmu}
\ee
with the two key constants
\bb
x \equiv \frac{3\sqrt{3}}{4} \approx 1.299\,038\,105\,677,
\label{x}
\ee
\bb
y \equiv \frac{15\pi}{64} \approx 0.736\,310\,778\,185.
\label{y}
\ee

(3) As effectively shown by Charles G.\ Darwin \cite{Darwin} for large deflection angle $\mu$ or small $\alpha = e^{-\mu}$,
\bb
\beta(\alpha) = 1 - z\alpha + O(\alpha^2)
\label{betasmallalpha}
\ee
with the third key constant
\bb
z \equiv \frac{216e^{-\pi}}{7+4\sqrt{3}} \approx 0.670\,165\,863\,511.
\label{z}
\ee

\section{The Order-[2,2] Two-Point Pad\'{e} Approximants}

Matching the behavior of the Pad\'{e} approximants Eqs.\ (\ref{betaa}) and (\ref{alphaa}) to the behavior of these results for $\beta(\alpha)$ near the two points $\alpha = 0$ and $\alpha = 1$ gives
\bb
A = \frac{4-4x+4y-2xz}{2x-2y-1} \approx 0.063\,359\,166\,591,
\label{A}
\ee
\bb
B = \frac{4-4x+4y-z-2yz}{2x-2y-1} \approx 0.733\,525\,030\,102,
\label{B}
\ee 
\bb
C = \frac{2-3x-2xy-xz+2xyz+2x^2}{x(2x-2y-1)} \approx -0.145\,150\,403\,080,
\label{C}
\ee
\bb
F = \frac{-0.5-x-y+2x^2z}{x(1-xz)} \approx -1.626\,918\,875\,112,
\label{F}
\ee
\bb
G = \frac{0.5-x-y-xz+2x^2z}{x(1-xz)} \approx -0.857\,118\,516\,193,
\label{G}
\ee
\bb
H = \frac{-0.5+y-0.5z-x^2z+2xz-yz}{x(1-xz)} \approx 0.107\,144\,750\,413.
\label{H}
\ee

These constant parameters then give the main result of this paper, the order-[2,2] two-point Pad\'{e} approximant Eq.\ (\ref{betaa}) for $\beta \equiv 3\sqrt{3}M/b \equiv$ (critical impact parameter)/(actual impact parameter) as a function of $\alpha \equiv e^{-\mu} \equiv$ the exponential of the negative of the light deflection angle $\mu$, that is, $\beta(\alpha) \approx \beta_p(\alpha)$, and vice versa by the order-[2,2] two-point Pad\'{e} approximant Eq.\ (\ref{alphaa}) for $\alpha(\beta) \approx \alpha_p(\beta)$, with the constants below rounded to 12 digits after the decimal point:
\bb
\beta_p(\alpha) \approx
\frac{1 + 0.063\,359\,166\,591\,\alpha - 1.063\,359\,166\,591\,\alpha^2}
{1 + 0.733\,525\,030\,102\,\alpha - 0.145\,150\,403\,080\,\alpha^2},
\label{betaaa}
\ee
and
\bb
\alpha_p(\beta) \approx 
\frac{1 - 1.626\,918\,875\,112\,\beta + 0.626\,918\,875\,112\,\beta^2}
{1 - 0.857\,118\,516\,193\,\beta + 0.107\,144\,750\,413\,\beta^2}.
\label{alphaaa}
\ee

\section{The Error of the Direct Pad\'{e} Approximants}

Obviously, $\alpha_p(\beta)$ is not the exact inverse function of $\beta_p(\alpha)$ (which would have square roots as the solution of a quadratic equation, but for 
$\beta_p(\alpha)$ and $\alpha_p(\beta)$ I am sticking to rational-function approximations), but within the allowed ranges $0 \leq\ \alpha \leq 1$ and $0 \leq\ \beta \leq 1$, $\alpha_p(\beta_p(\alpha))-\alpha$ (which is always nonnegative) has an rms value 0.000\,294 and a maximum value 0.000\,517, and $\beta_p(\alpha_p(\beta))-\beta$ (which is always nonpositive) has an rms value 0.000\,255 and a minimum value $-0.000\,460$, so for most practical purposes, these tiny discrepancies can be ignored.

In comparison with the exact $\alpha(\beta)$, given in terms of the elliptic integral in Eq.\ (\ref{alpha}), and its exact inverse $\beta(\alpha)$, the errors of the Pad\'{e} approximants $\alpha_p(\beta)$ and $\beta_p(\alpha)$ are slightly larger.  For example, $\alpha_p(\beta)-\alpha(\beta)$ is 0 at the endpoints $\beta = 0$ and at $\beta = 1$ (by design) and is slightly positive in between, with a maximum value of $0.002\,169$ at $\beta = 0.799\,627$, and with an rms error of $0.001\,168$.  Analogously, $\beta_p(\alpha)-\beta(\alpha)$ is also designed to have zero at the endpoints $\alpha = 0$ and $\alpha = 1$ and is also slightly positive in between, with a maximum value of $0.001\,433$ at $\alpha = 0.276\,807$, and with a root-mean-square (rms) error of $0.000\,840$.  (The small maximum errors of $\alpha_p(\beta)-\alpha(\beta)$ and of $\beta_p(\alpha)-\beta(\alpha)$ are very slightly different, mainly since although the exact $\beta(\alpha)$ is the inverse function of $\alpha(\beta)$, the order-[2,2] two-point Pad\'{e} approximants $\alpha_p(\beta)$ and $\beta_p(\alpha)$, both being rational functions with quadratic polynomials in both numerators and denominators, are not exact inverses of each other.)

\section{Simpler Quadratic Approximation}

Before giving the (often somewhat larger) errors of various functions of the Pad\'{e} approximants $\alpha_p$ (such as the approximate deflection angle $\mu_p = -\ln{\alpha_p}$) and of $\beta_p$ (such as the dimensionless approximate impact parameter $b_p/M = 3\sqrt{3}/\beta_p$), I want to give a quadratic approximation $\beta_q(\alpha) = 1-\alpha + K\alpha(1-\alpha)$, of the form so that $\beta_q(0) = 1$ and $\beta_q(1) = 0$ are exact, and for a suitable constant $K$.  To fit exactly
Einstein's small-angle deflection formula Eq.\ (\ref{muverysmallbeta}) that gives $\beta(\alpha) = x(1-\alpha) + O[(1-\alpha)^2]$ for small $1-\alpha$ with $x \equiv 3\sqrt{3}/4$, we would need $K$ to be
\bb
K = x-1 \equiv \frac{3\sqrt{3}}{4}-1
= \frac{3}{10}-\frac{1}{1040}-\frac{1}{1\,405\,040+811\,200\sqrt{3}} \approx 0.299\,038\,105\,677.
\label{K}
\ee
The denominator of the second term of the penultimate expression for $K$ has a form that might have been painfully familiar this month.
This value of $K$ then leads to
\bb
\beta_q(\alpha) = 1 - (2-x) \alpha - (x-1) \alpha^2
\approx 1 - 0.700\,961\,894\,323 \alpha - 0.299\,038\,105\,677 \alpha^2,
\label{betaq}
\ee
whose exact inverse function is
\bb
\alpha_q(\beta) = \frac{\sqrt{(x/2)^2-(x-1)\beta}-1+x/2}{x-1}.
\label{alphaq}
\ee
A simpler further approximation that is reasonably close to $\beta_q(\alpha)$ is
\bb
\beta_q(\alpha) \approx \beta_s(\alpha) = 1 - 0.7 \alpha - 0.3 \alpha^2,
\label{betaqapp}
\ee
whose inverse function is
\bb
\alpha_q(\beta) \approx \alpha_s(\beta) = \frac{\sqrt{169-120\,\beta}-7}{6}.
\label{alphaqapp}
\ee

\begin{figure}
\includegraphics[width=1.0\columnwidth]{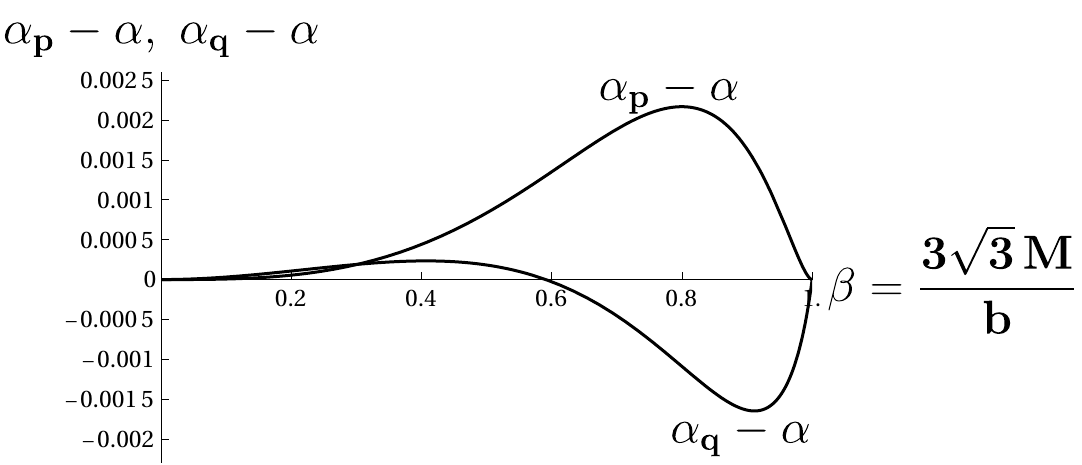}
\caption{This graph, with a greatly enlarged vertical scale from those of the previous two graphs, gives the error or deviation of the two-point Pad\'{e} approximant $\alpha_p(\beta)$ given by Eq.\ (\ref{alphaa}), and of the inverse quadratic approximation $\alpha_q(\beta)$ given by Eq.\ (\ref{alphaq}), away from the exact $\alpha(\beta)$ given by Eq.\ (\ref{alpha}).  
Note that the quadratic approximation $\alpha_q(\beta)$ has less error than the Pad\'{e} approximant $\alpha_p(\beta)$ over most of the range for $\beta$, but near the endpoints the Pad\'{e} approximant is more accurate.}
\end{figure}

\begin{figure}
\includegraphics[width=1.0\columnwidth]{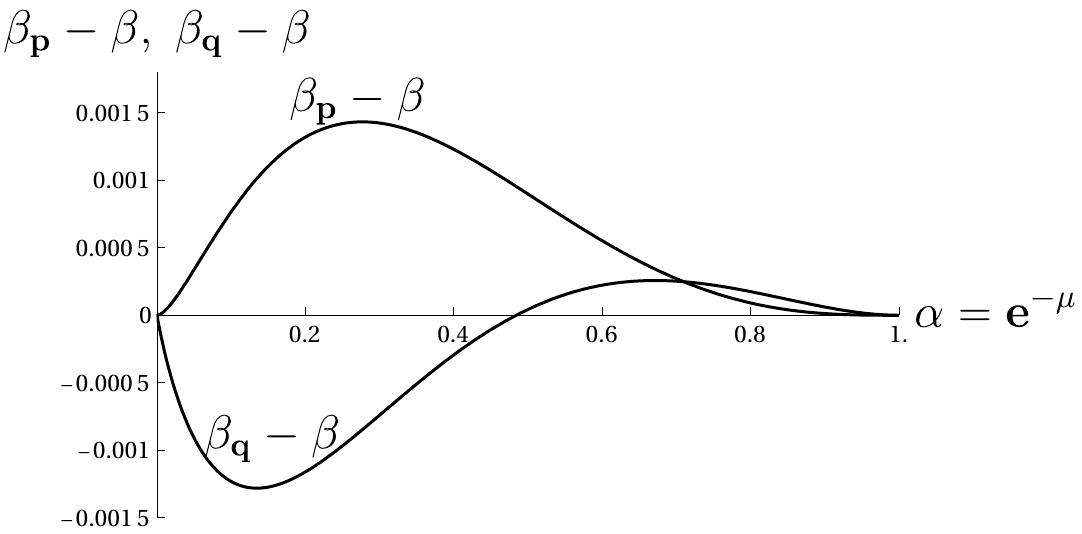}
\caption{This graph, also with a greatly enlarged vertical scale from those of the first two graphs, gives the error or deviation of the two-point Pad\'{e} approximant $\beta_p(\alpha)$ given by Eq.\ (\ref{betaa}), and of the quadratic approximation $\beta_q(\alpha)$ given by Eq.\ (\ref{betaq}), away from the exact $\beta(\alpha)$ given by the inverse of Eq.\ (\ref{alpha}).  
Note that the quadratic approximation $\beta_q(\alpha)$ has less error than the Pad\'{e} approximant $\beta_p(\alpha)$ over most of the range for $\alpha$, but near the endpoints the Pad\'{e} approximant is more accurate.}
\end{figure}

Unlike $\alpha_p(\beta)-\alpha(\beta)$ and $\beta_p(\alpha)-\beta(\alpha)$, which are always nonnegative and have maxima of $0.002\,169$ and $0.001\,433$ respectively for these Pad\'{e} approximants, the quadratic approximation $\beta_q(\alpha)$ and its inverse $\alpha_q(\beta)$ have $\beta_q(\alpha)-\beta(\alpha)$ and $\alpha_q(\beta)-\alpha(\beta)$ both switching sign in the interior of their domains.  For example, $\alpha_q(\beta)-\alpha(\beta) > 0$ for $0< \beta < 0.590\,693$ with a maximum of $0.000\,234$ at $\beta = 0.406\,681$, and then $\alpha_q(\beta)-\alpha(\beta) < 0$ for $0.590\,693 < \beta < 1$ with a minimum of $-0.001\,649$ at $\beta = 0.911\,373$.  On the other hand, 
$\beta_q(\alpha)-\beta(\alpha) < 0$ for $0 < \alpha < 0.590\,693$ with a minimum of $-0.001\,281$ at $\alpha = 0.134\,622$, and then $\beta_q(\alpha)-\beta(\alpha) > 0$ for $0.590\,693 < \alpha < 1$ with a maximum of $0.000\,257$ at $\alpha = 0.671\,154$.  Whereas the rms values of $\alpha_p(\beta)-\alpha(\beta)$ and $\beta_p(\alpha)-\beta(\alpha)$ are $0.001\,168$ and $0.000\,840$ respectively, the rms values of $\alpha_q(\beta)-\alpha(\beta)$ and $\beta_q(\alpha)-\beta(\alpha)$ are $0.000\,679$ and $000\,611$ respectively.

Thus these measures of the errors are smaller for the simpler quadratic approximation $\beta_q(\alpha)$ and its inverse $\alpha_q(\beta)$ given by Eqs.\ (\ref{betaq}) and (\ref{alphaq}) than for the order-[2,2] two-point Pad\'{e} approximants $\beta_p(\alpha)$ and $\alpha_p(\beta)$ (not quite the inverses of each other) given by Eqs.\ (\ref{betaa}) and (\ref{alphaa}) or by their truncated decimal forms Eqs.\ (\ref{betaaa}) and (\ref{alphaaa}).  However, the Pad\'{e} approximants have less error near the ends of the range for $\alpha = e^{-\mu}$ and $\beta = 3\sqrt{3}M/b$, e.g., for deflection angles $\mu = -\ln{\alpha}$ that are either very small or very large, corresponding to impact parameters $b = 3\sqrt{3}M/\beta$ either very much larger than the black hole mass $M$ or near the critical value $3\sqrt{3}M$, the minimum value for the impact parameter $b$ for an idealized point photon not to fall into the black hole.  Therefore, we now turn to measures of the errors for other functions of $\alpha$ and $\beta$.

\section{Several Pad\'{e} and Quadratic Error Measures}

Although surprisingly the quadratic approximation has less error for $\beta(\alpha)$ and $\alpha(\beta)$ than the Pad\'{e} approximations when averaged over the unit range for either $\alpha$ or $\beta$, the Pad\'{e} approximations are more accurate near the end of either of these ranges for functions of $\alpha$ and of $\beta$ that diverge at the corresponding endpoints.  For example, as $\alpha = e^{-\mu}$ approaches 1, the deflection angle $\mu = -\ln{\alpha}$ approaches 0, and hence $\beta(\alpha)$ approaches 0, so that then the impact parameter $b = 3\sqrt{3}M/\beta$ grows without bound (diverges to infinity).  At the other extreme ($\beta$ approaching 1), the impact parameter $b=3\sqrt{3}M/\beta$ approaches the critical value $b_c = 3\sqrt{3}M)$, $\alpha$ approaches 0, and the deflection angle $\mu = -\ln{\alpha}$ grows without bound, as the idealized point photon orbits around the black hole an arbitrarily large number of times as it asymptotically approaches the circular photon orbit at $r=3M$.  Therefore, for functions that diverge at an endpoint (such as the impact parameter $b=3\sqrt{3}M/\beta$ at $\alpha = 1$ and thus at $\beta = 0$, or the deflection angle $\mu = -\ln{\alpha}$ at $\beta = 1$ and thus at $\alpha = 0$), the maximum and rms errors are generally larger for the quadratic approximation than for the Pad\'{e} approximations.


\begin{table}
\begin{tabular}{||l|l|l|l|l||} \hline
Function & Pad\'{e} Max & Quad Max & Pad\'{e} RMS & Quad RMS \\ \hline
$\alpha_a - \alpha$ & 0.002\,169 & 0.001\,649 & 0.001\,168 & 0.000\,679 \\ \hline
$\frac{\alpha_a}{\alpha}-1$ & 0.012\,386 & 0.043\,934 & 0.005\,399 & 0.008\,057 \\ \hline
$\frac{\alpha_a - \alpha}{1-\alpha}$ & 0.003\,050 & 0.001\,899 & 0.001\,844 & 0.000\,925 \\ \hline
$\mu_a-\mu$ & 0.012\,310 & 0.044\,928 & 0.005\,373 & 0.008\,166  \\ \hline
$\frac{\mu_a}{\mu}-1$ & 0.006\,361 & 0.008\,721 & 0.003\,576 & 0.002\,966 \\ \hline 
$\beta_a - \beta$ & 0.001\,433 & 0.001\,281 & 0.000\,840 & 0.000\,611 \\ \hline
$\frac{\beta_a}{\beta}-1$ & 0.001\,887 & 0.001\,432 & 0.001\,201 & 0.000\,793 \\ \hline
$\frac{\beta_a-\beta}{1-\beta}$ & 0.010\,687 & 0.045\,953 & 0.005\,182 & 0.009\,974 \\ \hline
$\frac{1}{\beta_a}-\frac{1}{\beta}$ & 0.002\,749 & 0.004\,704 & 0.001\,950 & 0.002\,105 \\ \hline
\end{tabular}
\caption{
Here the first column gives the function denoting the type of error, with the subscript $a$ (for ``approximate'') representing either the Pad\'{e} or the quadratic approximation, the second and third columns give the maximum absolute values for these errors, and the fourth and fifth columns give the root-mean-square deviations of these errors from zero when averaged over the unit range of the corresponding function argument $\beta$ or $\alpha$.  Note that for these functions, the maximum error of the Pad\'{e} approximations is less than one part in 80 (1.25\%), whereas the maximum error of the quadratic approximations is slightly more than one part in 22 (nearly 4.6\%).  However, for functions that do not diverge (as the impact parameter $b=3\sqrt{3}M/\beta$ does at zero deflection angle $\mu$ where $\alpha \equiv e^{-\mu} = 1$, and as the deflection angle $\mu = -\ln{\alpha}$ does at the critical impact parameter $b = b_c = 3\sqrt{3}M$ where $\beta \equiv 3\sqrt{3}M/b = 1$), the simpler quadratic approximations often have less error.
}
\end{table}

\begin{figure}[H]
\includegraphics[width=1.0\columnwidth]{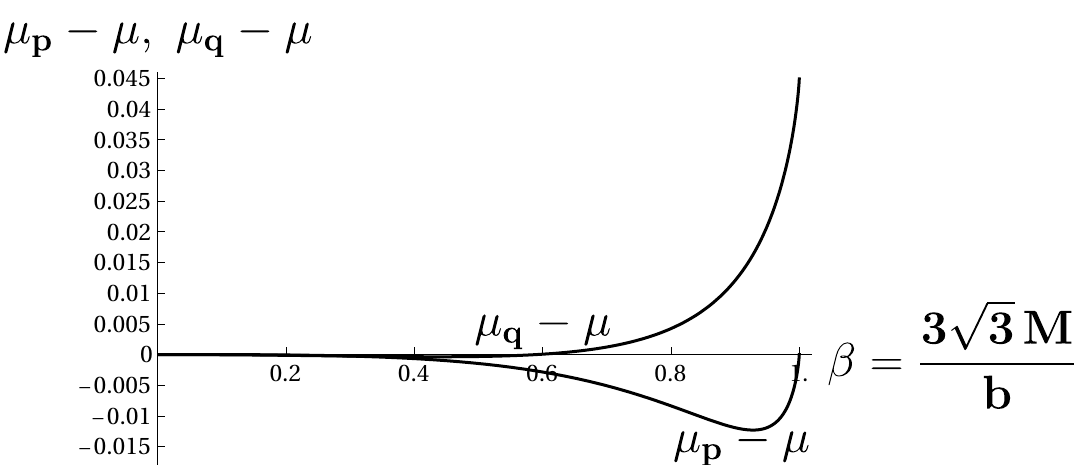}
\caption{This graph, with a significantly enlarged vertical scale from those of the previous two graphs, gives the error or deviation of $\mu_p(\beta) = -\ln{\alpha_p}$ with the two-point Pad\'{e} approximant $\alpha_p(\beta)$ given by Eq.\ (\ref{alphaa}), and of $\mu_q(\beta) = -\ln{\alpha_q}$ with the inverse quadratic approximation $\alpha_q(\beta)$ given by Eq.\ (\ref{alphaq}), away from the exact $\mu(\beta) = -\ln{\alpha}$ with $\alpha(\beta)$ given by Eq.\ (\ref{alpha}).
}
\end{figure}

\begin{figure}[H]
\includegraphics[width=1.0\columnwidth]{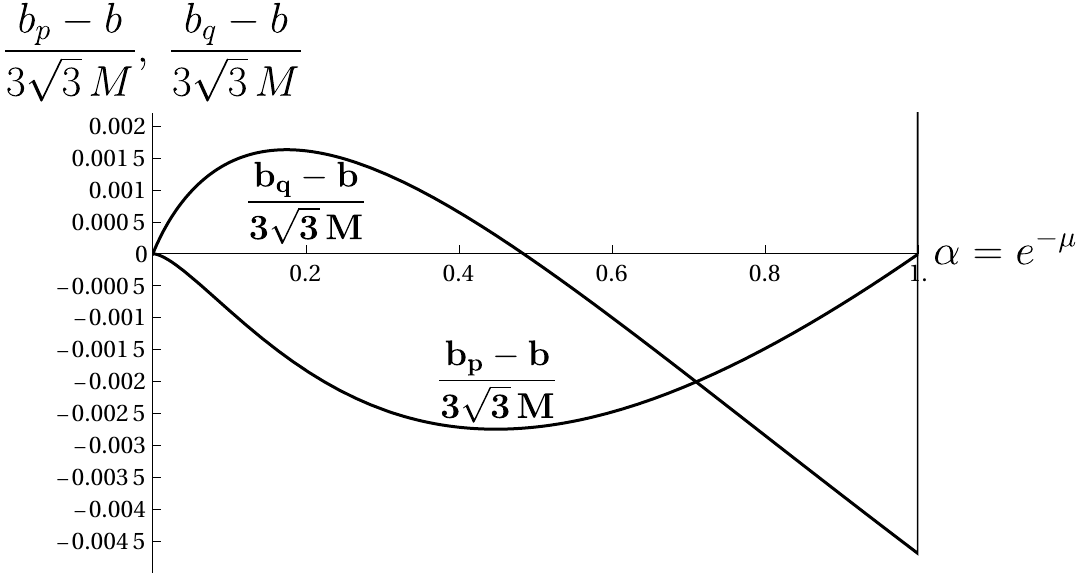}
\caption{This graph, also with a significantly enlarged vertical scale from those of the first two graphs, gives
$\frac{b_p-b}{3\sqrt{3}\,M}$
and $\frac{b_q-b}{3\sqrt{3}\,M}$,
the error or deviation (from the exact normalized impact parameter
$\frac{b}{3\sqrt{3}\,M} = \frac{1}{\beta(\alpha)}$)
of the normalized impact parameter
$\frac{b_p}{3\sqrt{3}\,M} = \frac{1}{\beta_p(\alpha)}$ from the
two-point Pad\'{e} approximant $\beta_p(\alpha)$ in Eq.\ (\ref{betaa}), 
and the error of the normalized impact parameter
$\frac{b_q}{3\sqrt{3}\,M} = \frac{1}{\beta_q(\alpha)}$ from
the quadratic approximation $\beta_q(\alpha)$ in Eq.\ (\ref{betaq}).
}
\end{figure}

\newpage

\section{Conclusions}

In the Schwarzschild metric for a nonrotating black hole, a light ray of impact parameter $b = 3\sqrt{3}M/\beta$ has a deflection angle $\mu = \ln{\alpha}$ given exactly by Charles G.\ Darwin \cite{Darwin} in terms of an elliptic integral, as summarized in the Introduction.  Here it is shown that the functional relation between the dimensionless parameters $\beta \equiv 3\sqrt{3}M/b$ and $\alpha \equiv e^{-\mu}$ may be well approximated by rational functions.  Excellent fits over the full unit range of both $\beta(\alpha)$ and $\alpha(\beta)$ are the order-[2,2] Pad\'{e} 2-point approximant $\beta_p(\alpha)$ in Eq.\ (\ref{betaa}) with its three constants given exactly in Eqs.\ (\ref{A})-(\ref{C}) and approximately in Eq.\ (\ref{betaaa}), and the corresponding order-[2,2] Pad\'{e} 2-point approximant 
$\alpha_p(\beta)$ in Eq.\ (\ref{alphaa}) with its three constants given exactly in Eqs.\ (\ref{F})-(\ref{H}) and approximately in Eq.\ (\ref{alphaaa}) for the inverse function.  A simpler fit for $\beta(\alpha)$ is the quadratic approximation $\beta_q(\alpha)$ of Eq.\ (\ref{betaq}) and its inverse $\alpha_q(\beta)$ of Eq.\ (\ref{alphaq}).  An even simpler fit is the 1-digit approximation to the constant in Eq.\ (\ref{betaq}) that gives Eq.\ (\ref{betaqapp}), $\beta_s(\alpha) = 1-0.7\alpha-0.3\alpha^2$.

As elementary functions approximating $\beta(\alpha)$ and $\alpha(\beta)$, all of these approximations have errors for these quantities always less than 1 part in 460.  However, when one considers, as a function of the impact parameter $b$ or its dimensionless inverse impact parameter $\beta = 3\sqrt{3}M/b$, the deflection angle $\mu = \ln{\alpha}$, which diverges in the limit $\beta \rightarrow 1$, has its Pad\'{e} approximation with a maximum absolute error of nearly one part in 81, whereas the quadratic approximation has its maximum absolute error of nearly one part in 22, about 3.6 times larger.  Although the quadratic approximations surprisingly work as well as the Pad\'{e} approximations over most of the unit range of $\beta$ and $\alpha$, when one goes to impact parameters $b$ near the critical impact parameter $b_c = 3\sqrt{3}M$ or $\beta = b_c/b$ near 1, the Pad\'{e} approximation gives much less error for the deflection angle $\mu = \ln{\alpha}$ (absolute error going to zero in the limit) than the quadratic approximations (which give nonzero absolute errors in the limit, though only about
$2.6^\circ$).  Similarly, when one goes to very small deflection angles $\mu$ or $\alpha = e^{-\mu}$ near 1, where the impact parameter $b = 3\sqrt{3}M/\beta$ is very large, the Pad\'{e} approximation gives an absolute error for the impact parameter that goes to zero in the limit of infinite impact parameter, whereas the quadratic approximations give absolute errors approaching $-0.024M$.

\section*{Acknowlegments}

This research was supported in part by the Natural Sciences and Engineering Research Council of Canada (NSERC).

\end{document}